\documentclass[twocolumn,groupedaddress,amsmath,amssymb,prl,aps]{revtex4}

\usepackage{graphicx}
\usepackage{dcolumn}
\usepackage{bm}
\usepackage{soul} 
\usepackage{xcolor} 
\usepackage{amsmath} 

\begin{document}

\title{Elastic amplification from negative capacitance}

\author{M\'onica Graf,$^{1,2}$ Natalya S. Fedorova,$^{1}$ Hugo
  Aramberri,$^{1}$ and Jorge \'I\~niguez-Gonz\'alez$^{1,3}$}

\affiliation{
  \mbox{$^{1}$Luxembourg Institute of Science and Technology (LIST), Avenue des Hauts-Fourneaux 5, L-4362 Esch/Alzette,     Luxembourg}\\
  \mbox{$^{2}$Department of Dielectrics, Institute of Physics of the Czech Academy of Sciences (FZU-AVCR),} \mbox{Na Slovance 1999/2, 182 00 Prague, Czechia}\\
 \mbox{$^{3}$Department of Physics and Materials Science, University of Luxembourg, Rue du Brill 41, L-4422 Belvaux, Luxembourg}}

\begin{abstract}
    Ferroelectrics under suitable electric boundary conditions can present a negative capacitance response, whereby the total voltage drop across the ferroelectric opposes the externally applied bias. When the ferroelectric is in a heterostructure, this behavior yields a voltage amplification in the other elements, an effect that could be leveraged in low-power electronic devices. Interestingly, the mentioned voltage amplification should have an accompanying elastic effect. Specifically, in the typical case that the materials in contact with the ferroelectric are non-polar dielectrics, those should present an enhanced electrostrictive response. Here we use atomistic simulations -- of model PbTiO$_{3}$/SrTiO$_{3}$ ferroelectric/dielectric superlattices displaying negative capacitance -- to show that this is indeed the case: we reveal the enhanced elastic response of the dielectric layer and show that it is clearly linked to the voltage amplification. We argue that this ``elastic amplification'' could serve as a convenient experimental fingerprint for negative capacitance. Further, we propose that it may have interest on its own, e.g. for the development of low-power electromechanical actuators.
\end{abstract}

\maketitle   

A ferroelectric material under suitable electric boundary conditions can display steady-state negative capacitance (NC)~\cite{zubko16}. In ferroelectric/dielectric heterostructures, for instance, the dielectric precludes the full development of the polarization in the ferroelectric. Consequently, the latter will remain in an electrostatically frustrated state -- either in a multidomain configuration or as an ``incipient ferroelectric'' -- with a higher energy than the single-domain state featuring a fully developed polarization. Then, upon the application of an external electric field, the ferroelectric can access states with a finite polarization, allowing it to decrease its energy (instead of increasing it like a normal capacitor would). Therefore, the ferroelectric behaves as a negative capacitance~\cite{iniguez19,bratkovsky01,bratkovsky06}. The energy released by the ferroelectric is effectively transferred to the dielectric, which feels an electric field higher than the one applied, i.e. we have a ``voltage amplification'' (VA). Indeed, this constitutes the most promising application of negative capacitance, as it can lead to low-power consuming devices.

Unfortunately, it is not easy to measure this effect. For instance, measured capacitance enhancements in ferrolectric/dielectric bilayers -- or similar structures -- are usually taken as a signature of negative capacitance in the ferroelectric~\cite{khan11,appleby14}. Yet, other effects -- from Maxwell-Wagner currents \cite{catalan00} to features of the ferroelectric/electrode interface \cite{stengel06} -- could also explain the measured capacitance increases, which often precludes a definite cause-effect connection~\cite{iniguez19}. Therefore, it would be convenient to find alternative ways to identify the appearance of this property in a measurable, experimental manner.

Interestingly, all insulating materials subject to an electric voltage display an elastic response known as electrostriction, that is, a variation of the strain that is quadratic in the applied field~\cite{newnham97}. It follows that, if we have a VA in the dielectric, we should also have a similarly amplified electrostrictive response. Note that VA is clearly linked to an underlying NC effect, as the amplification is {\sl powered} by the energy released by the electrically-frustrated ferroelectric. Further, let us stress that the spurious effects mentioned above (e.g., Maxwell-Wagner currents) cannot produce VA or an enhanced elastic response. Hence, an amplified elastic response is a direct signature of VA, which in turn can be safely attributed to NC behavior.

In this work we demonstrate that this is indeed the case, by presenting a proof-of-concept theoretical study of ferroelectric/dielectric superlattices displaying NC. We see that, when the ferroelectric layer is in the NC regime, the strain response of the dielectric layers is enhanced, and that such an enhancement is unequivocally linked to the NC effect. This ``elastic amplification'' may serve as a fingerprint of VA and NC. Further, it may also have an applied interest by itself, as it yields an enhanced (low-power) electromechanical actuation.

In order to investigate the effect that NC has on the electrostrictive response, we study PbTiO$_3$/SrTiO$_3$ superlattices by means of second-principles simulations~\cite{wojdel13,scaleup,zubko16}. This approach has been successfully applied to both bulk materials~\cite{wojdel13,graf21,murillo21} and ferroelectric/dielectric heterostructures~\cite{zubko16,aramberri22,graf22}, to explain a wide variety of properties ranging from caloric \cite{murillo23} and thermal \cite{seijas-bellido17} effects to NC responses \cite{zubko16,graf22} or the onset of polar skyrmions~\cite{das19,goncalves19}.

We work with superlattices that have been recently predicted to display giant voltage amplifications~\cite{graf22}. More precisely, we consider 9/6 heterostructures composed of PbTiO$_{3}$ (PTO) layers that are 9 unit cells thick and SrTiO$_{3}$ (STO) layers that are 6 unit cells thick. We also consider an epitaxial tensile strain of 1~\% with respect to STO's bulk lattice parameter ($a_{0} = 3.901$~\AA), which is close to the strain obtained in STO films grown on (110) DyScO$_3$ substrates~\cite{tyunina09}. The external field ($\mathcal{E}_{\rm {ext}}$) is applied along the superlattice stacking direction -- i.e., the $z$ Cartesian direction in the following -- ranging between $-$1.5~MV/cm and 1.5~MV/cm. Also, for simplicity and computational convenience, we work at 0~K, noting that the results should be representative of any regime where NC is present. See Supplementary Note~1 for more details.

We modify our second-principles superlattice model in a way that allows us to vary continuously (and eventually turn off) the NC response of the PTO layer, so we can monitor the effect on the elastic response of the STO layer. We do this by controlling the response properties of the PTO layer, making it electrically stiffer or softer. As detailed in Ref.~\cite{wojdel13}, our second-principles approach gives us precise control of the interatomic interactions. In particular, here we add a harmonic spring constant between Ti--O and Pb--O neighboring atom pairs, which can be expressed as
\begin{equation}
\begin{split}
  \Delta E = & \lambda \left[ \sum_{i} \sum_{j \/ i} \sum_{\alpha} ( u_{{\rm Ti},i\alpha} - u_{{\rm O},j\alpha})^{2} 
    \right. \\
    & \left. + \sum_{i} \sum_{j \/ i} \sum_{\alpha} (u_{{\rm Pb},i\alpha} - u_{{\rm O},j\alpha})^{2} \right] \; .
\end{split}
\end{equation}
Here, $u$ denotes atomic displacements with respect to the ideal (cubic-like) perovskite structure, used as reference to construct the second-principles model~\cite{wojdel13}; $\alpha$ labels Cartesian directions; the $i$ index runs over all the elemental 5-atom perovskite cells in the PTO layer (hence, the two sums include all Ti and Pb atoms in the ferroelectric layer, respectively), while the $j$ index runs over all oxygens that are nearest neighbors of the $i$-th Ti or Pb. Finally, $\lambda$ is a stiffness constant that has units of meV/\AA$^2$ and which we can tune at will.

Note that when $\lambda > 0$ we penalize the atomic displacements that lead to the development of an electric polarization in PTO, while we favor them when it is negative. Thus, we can make PTO less (or more) responsive to external electric stimuli. Experimentally, this could be achieved by doping the PTO layer with other chemical species that modify PTO's polarizability and tendency to be ferroelectric (see Supplementary Figure~2). Also, note that $\lambda = 0$ corresponds to the unmodified superlattice, which is in the NC regime with an estimated inverse electric permittivity of $-$0.019~$\epsilon_{0}^{-1}$ for the PTO layer and a voltage amplification ratio of 4.7 in the STO layer. (These values were obtained (at 0~K) following the methodology described in Ref.~\cite{graf22} and summarized in Supplementary Note~2.) Finally, let us stress that we only modify the interactions within the PTO layer, leaving STO untouched.

\begin{figure}[th!]
    \centering
    \includegraphics[width=\columnwidth]{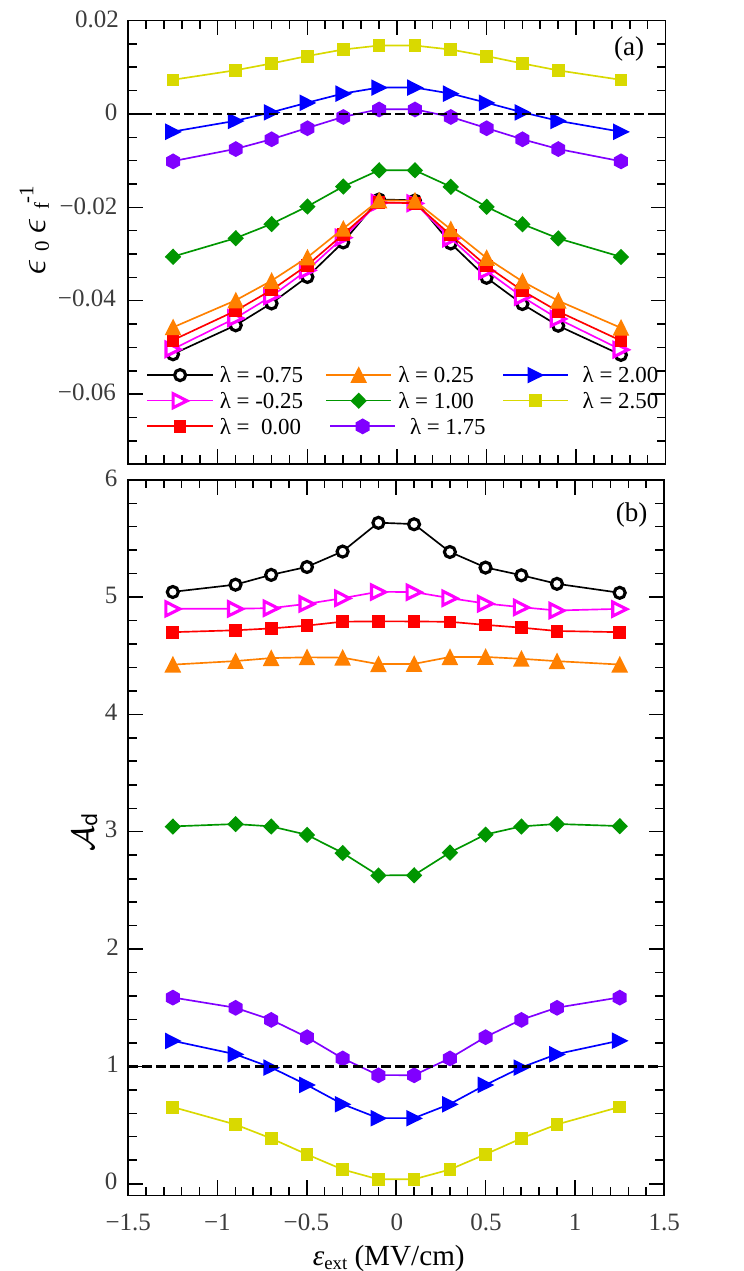}
    \caption{Inverse electric permittivity, $\epsilon_{0}\epsilon_{\rm f}^{-1}$, of the PTO layer (a) and voltage ratio, $\mathcal{A}_{\rm d}$, in the STO layer (b) as functions of the external electric field onto the (PTO)$_{9}$/(STO)$_{6}$ superlattices with different stiffness parameters $\lambda$ given in meV/\AA$^{2}$.}
    \label{fig:permit}
\end{figure}

Having defined our model superlattices, we first analyze how $\lambda$ affects their NC and VA behavior. Our results are summarized in Fig.~\ref{fig:permit}, which displays the calculated permittivity of the ferroelectric layer ($\epsilon_{0}\epsilon_{\rm f}^{-1}$) in panel~(a) and the differential VA in the STO layer ($\mathcal{A}_{\rm d} = d V_{\rm d}/ d V_{\rm ext}$) in panel~(b). The NC regime corresponds to having $\epsilon_{0}\epsilon_{\rm f}^{-1}<0$ and $\mathcal{A}_{\rm d} >1$. Hence, for instance, we see that for soft PTO layers -- i.e., for $\lambda \leq 1$ meV/\AA$^{2}$ -- we obtain NC, with maximum VA for the superlattice with the softest (most responsive) PTO layer considered -- i.e. $\lambda = -0.75$ meV/\AA$^2$. Then, as $\lambda$ takes higher positive values, the superlattice progressively moves towards a normal capacitance regime where the permittivity of the PTO layer becomes positive and, consequently, the (differential) voltage drop in the STO layer is smaller than the external voltage applied ($\mathcal{A}_{\rm d}<1$). We thus confirm that less responsive ferroelectric layers do not present NC, as we expected. 

From these results, we also note that -- in most cases -- the permittivity of the PTO layer remains negative in the whole range of applied electric fields, which suggests that the NC effect is robust against the application of sizeable biases. (In our theoretical calculations, an external field of 1~MV/cm corresponds to a voltage drop of about 0.6~V across one superlattice period.) Also remarkable is the fact that, in some cases, the superlattice enters the NC regime for high enough external fields, which reflects an underlying transformation of the PTO layer. These aspects  -- which concern the NC response in the non-linear regime -- go beyond the scope of the present work and will be addressed in a future publication.

\begin{figure}[th!]
    \centering
    \includegraphics[width=\columnwidth]{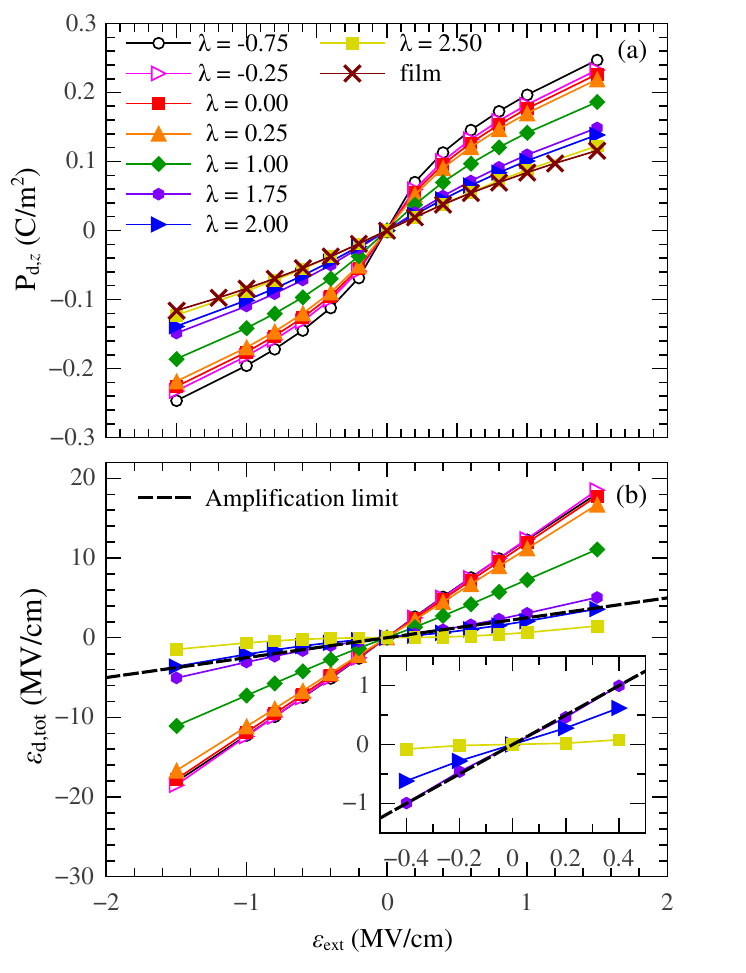}
    \caption{Polarization (a) and total field ($\mathcal{E}_{\rm d,tot}$) (b) of STO layer as functions of the external field. Panel (a) also shows the polarization of a STO film under the effect of the same external field for comparison. The inset in panel (b) shows the total field of STO layer for the highest stiffness coefficients at low external field.}
    \label{fig:pol_indField}
\end{figure}

We now focus on the dielectric response of the STO layer in the superlattice. Our results are summarized in Fig.~\ref{fig:pol_indField}; for comparison, we include the results for a STO film subject to the same elastic clamping assumed for the superlattices (see Supplementary Note~3 for details). Figure~\ref{fig:pol_indField}(a) shows the polarization of the STO layer ($P_{{\rm d},z}$) as a function of the applied external field. As expected, STO gets more strongly polarized as the VA increases, i.e., as $\lambda$ decreases. Let us note that, in the absence of VA (normal regime), the maximum total field that could be observed in the STO layer would correspond to the situation in which the PTO layer behaves as a perfect metal (infinitely polarizable) and the applied voltage falls exclusively in the dielectric. Hence, this maximum field is simply given by $\mathcal{E}_{\rm d}^{\rm max} = L \mathcal{E}_{\rm ext}/l_{\rm d}$, where $l_{\rm d}$ is the thickness of the dielectric STO layer and $L$ is the thickness of a full superlattice period. This maximum-field limit is marked with a dashed line in Fig.~\ref{fig:pol_indField}(b). For the softest ($\lambda < 1.75$~meV/\AA$^2$) and hardest ($\lambda > 2.00$~meV/\AA$^2$) PTO layers, respectively, the field in the STO layer clearly exceeds this maximum or falls short of it. Additionally, we have borderline cases (e.g., $\lambda = 2.00$~meV/\AA$^2$) where we have a normal behavior for small external fields (${\cal A}_{\rm d}<1$) and a sufficiently large bias is required to induce a differential amplification (${\cal A}_{\rm d}>1$). In such cases it is crucial to account for the non-linear response of the system, and we need to recall that the total field in the dielectric can be obtained from ${\cal A}_{\rm d}$ as
\begin{equation}
{\cal E}_{\rm d, tot} =  \int_{0}^{{\cal E}_{\rm ext}} \frac{L}{l_{\rm d}} {\cal A}_{\rm d}\, d{\cal E}'_{\rm ext} \approx \frac{L}{l_{\rm d}} \int_{0}^{{\cal E}_{\rm ext}} {\cal A}_{\rm d}\, d{\cal E}'_{\rm ext} \; ,
\end{equation}
where all the quantities in the integrand are a function of the external field and, for the sake of this qualitative argument, in the last step we assume that the dependence of the layer thicknesses is relatively weak. It is thus clear that, whenever we start with ${\cal A}_{\rm d}<1$ at small biases, we need to go well into the regime with differential VA in order to obtain an amplified field $|\mathcal{E}_{\rm d, tot}|> L |\mathcal{E}_{\rm ext}|/l_{\rm d}$.

\begin{figure*}[th!]
    \centering
    \includegraphics[width=16cm]{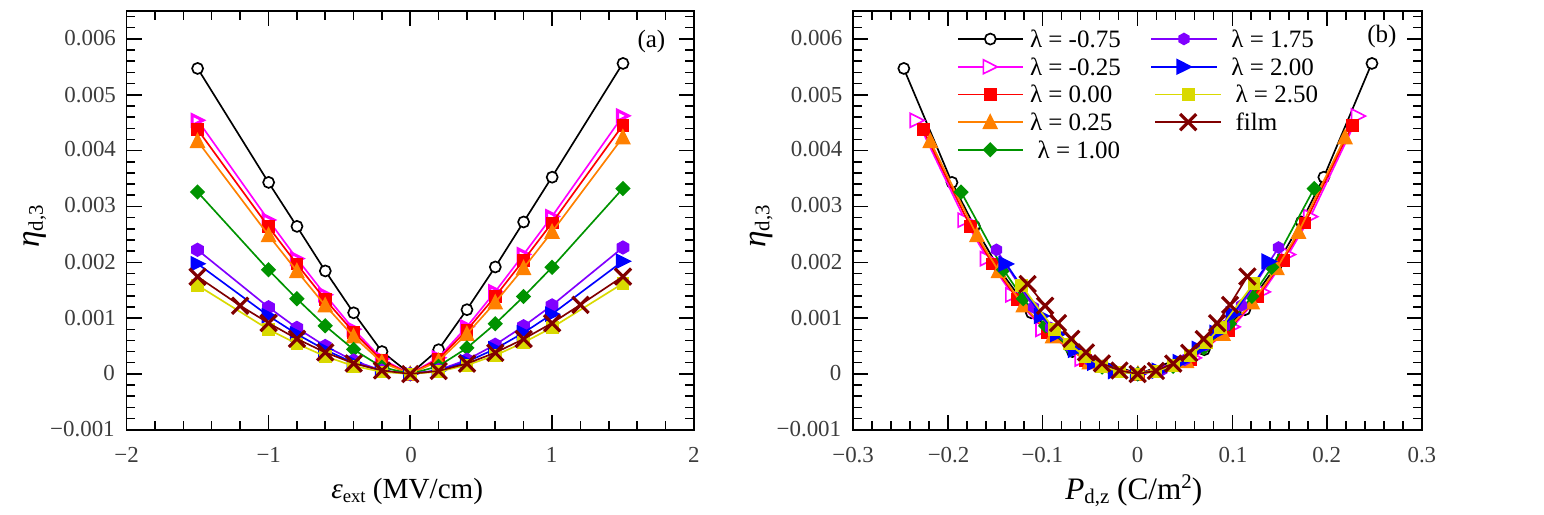}
    \caption{Out-of-plane elastic response of the STO layer and the STO film as a function of the external field (a) and the out-of-plane polarization (b) for the different choices of $\lambda$ parameter.}
    \label{fig:eta3}
\end{figure*}

Let us now turn our attention to the elastic response of the STO layer (i.e., the change of its thickness) as a function of the applied external field and for the different choices of $\lambda$. Figure~\ref{fig:eta3}(a) shows our results for the relevant strain component ($\eta_{{\rm d},3}$), including the case of the STO film for comparison. We find an essentially parabolic behavior, as expected for an electrostrictive effect. Further, we find that the maximum elastic response corresponds to $\lambda=-0.75$ meV/\AA$^{2}$, i.e., the case where we have the strongest VA. By contrast, as $\lambda$ grows, the STO layer elongates less and less. Note also that, whenever we are in the NC regime, the observed effect is stronger than the one obtained for a pure STO film. 

This clear connection between VA and the elastic response of the STO layer makes good physical sense. Note that the intrinsic response properties of the STO layer must be independent from $\lambda$, as this parameter only affects the interactions within the PTO layer. Further, beyond some short-range interactions between nearest-neighboring atoms at the interface, the PTO and STO layers are only coupled by electrostatic forces, whose main manifestation is captured by the VA ratio. Hence, our results suggest a simple picture whereby the NC effect yields an amplified electric field in the STO layer (Fig.~\ref{fig:pol_indField}(b)), which in turn results in an enhanced polarization (Fig.~\ref{fig:pol_indField}(a)), which finally results in a relatively large strain (Fig.~\ref{fig:eta3}(a)). From this picture, the electrostrictive coefficient $Q$, as defined by $\eta_{{\rm d},3} = Q P_{{\rm d},z}^{2}$, would be independent of $\lambda$. This is clearly verified in Fig.~\ref{fig:eta3}(b), which shows how the $\eta_{{\rm d},3}=\eta_{{\rm d},3}(P_{{\rm d},z})$ curves obtained for different $\lambda$ values collapse into one, and also coincide with the result for a pure STO film. From these curves we get $Q \approx 0.13$~m$^{4}$/C$^{2}$.

Additionally, from Fig.~\ref{fig:eta3}(a) we can estimate an effective electrostrictive coefficient $M^{\rm {eff}}$, defined as $\eta_{{\rm d},3} = M^{\rm eff} \mathcal{E}_{\rm ext}^{2}$. We obtain values between $2.8\times 10^{-19}$~m$^{2}$/V$^{2}$ (for $\lambda = -0.75$ meV/\AA$^{2}$, maximum VA) and $0.74 \times 10^{-19}$~m$^{2}$/V$^{2}$ (for $\lambda = 2.5$ meV/\AA$^{2}$, normal regime). For the pure STO film we get $0.82 \times 10^{-19}$~m$^{2}$/V$^{2}$, again consistent with the fact that, for the same applied external field, the electrostrictive effect in the dielectric layer is enhanced by the NC response of the ferroelectric one.

We have not found experimental data to compare directly with our estimated electrostrictive coefficients. Nevertheless, let us note that an electrostrictive coefficient of 0.088~m$^{4}$/C$^{2}$ was obtained through first-principles calculations of bulk STO in Ref.~\cite{jiang16}. (This coefficient corresponds to the longitudinal response to a field applied along the axis of the oxygen octahedra tilts present in STO.) Our second-principles prediction for the corresponding electrostrictive coefficient of bulk STO is 0.11~m$^{4}$/C$^{2}$, in reasonable agreement with the first-principles result. 

In conclusion, in this work we show that, when negative capacitance occurs in the ferroelectric layer of a heterostructure, it comes with two accompanying effects: a well-known electric one (i.e., a voltage amplification in the rest of the circuit) and a previously unnoticed elastic one (i.e., an enhanced elastic response -- typically, electrostrictive -- that is a direct consequence of the amplified voltage). We believe this ``elastic amplification'' is interesting for at least two  reasons. On one hand, it may provide us with a convenient and reliable experimental fingerprint of voltage amplification and negative capacitance. Indeed, while challenging, measuring the layer thickness -- and its response to an external bias -- seems more viable than monitoring the local electric fields. On the other hand, an amplified elastic response suggests the possibility of building low-power electromechanical actuators, which would yield a desired deformation at a reduced applied voltage. We thus hope our predictions will stimulate research on the mechanical consequences of negative-capacitance effects.

We thank Pavlo Zubko (UC London) for fruitful discussions. Work funded by the Luxembourg National Research Fund through grants INTER/RCUK/18/12601980 and C21/MS/15799044/FERRODYNAMICS.

\section{Methods}

The second-principles methods used in this work are described in detail in Refs.~\cite{wojdel13,garciafernandez16}. The models for the superlattices are derived from models for bulk SrTiO$_3$ and bulk PbTiO$_3$ that have been used in previous works and give accurate descriptions of the lattice dynamical properties of both compounds. 

The simulation supercell for the considered 9/6 PTO/STO supelattices can be seen as composed of 8$\times$8$\times$15 perovskite-like units, with periodic boundary conditions along the three spatial directions.

To simulate pure STO films, while keeping all other approximations identical as for the superlattice simulations, we constructed as a superlattice with very thick STO layers (30 perovskite-like cells) and ultra-thin PTO layers (only 1 cell). This thin ferroelectric layer does not result in any NC effect, as described in the Supplementary Note~3. In this case, the simulation supercell can be seen as composed of 2$\times$2$\times$31 perovskite-like cells with periodic boundary conditions.

We compute the strain associated to the dielectric layer as
\begin{equation}
    \eta_{{\rm d},3} = \frac{c_{\rm d}(\mathcal{E}_{\rm {ext}})-c_{\rm d}(0)}{c_{\rm d}(0)} \; ,
\end{equation}
where $c_{\rm d}(\mathcal{E}_{\rm {ext}})$ is the lattice parameter of the STO layer along the $z$ direction as obtained for an applied external electric field $\mathcal{E}_{\rm {ext}}$.

\end{document}